\documentclass [aps, prl, preprint, doublespace, showpacs, amsmath, floatfix, superscriptaddress]{revtex4-1}

\usepackage{graphicx}
\begin{document}

\title{Quasiparticle Excitations and Charge Transition Levels of Oxygen Vacancies in Hafnia}

\author{Manish Jain}
\affiliation{Department of Physics, University of California, Berkeley, California 94720}
\affiliation{Materials Sciences Division, Lawrence Berkeley, National Laboratory, Berkeley, California 94720}
\author{James R. Chelikowsky}
\affiliation{Center for Computational Materials, Institute for Computational Engineering and Sciences, Departments of Physics and Chemical Engineering, University of Texas, Austin, 78712}
\author{Steven G. Louie}
\affiliation{Department of Physics, University of California, Berkeley, California 94720}
\affiliation{Materials Sciences Division, Lawrence Berkeley, National Laboratory, Berkeley, California 94720}

\date{\today}

\begin{abstract}
We calculate the quasiparticle defect states and charge transition levels of 
oxygen vacancies in monoclinic hafnia. The charge transition levels, although 
they are thermodynamic quantities, can be critically dependent on the band gap 
owing to localized defect states. These quasiparticle defect level effects are 
treated using the first principle GW approximation to the self energy. We show 
that the quality and reliability of the results may be evaluated by calculating 
the same transition level via two physical paths and that it is important to 
include the necessary electrostatic corrections in a supercell calculation. 
Contrary to many previous reports, the oxygen vacancies in monoclinic hafnia are found 
to be a positive U center, where U is the defect electron addition energy. We 
identify a physical partitioning of U in terms of an electronic and structural 
relaxation part.  
\end{abstract}


\maketitle


Hafnia has recently received much attention because of its many applications, in 
particular as high-dielectric gate material replacing silica in microelectronic 
devices. However, devices based on hafnia suffer from several problems  such as 
voltage threshold instabilities \cite{Ribes05} and flat band voltage shifts 
\cite{Hobbs04}.  These problems are believed to be due to a high density of 
defects in the material  in particular oxygen vacancies are believed to play an 
important role as electron traps.

There have been several theoretical studies on the structural and electronic 
properties of oxygen vacancies in monoclinic hafnia. In early studies, formation 
energies as well as defect levels were calculated within density functional 
theory (DFT) using the local density approximation \cite{Foster02} and the 
generalized gradient approximation (GGA) \cite{Zheng07,*Kang03}. The onsite 
energy, U, for adding an additional electron to the defect was calculated and the 
vacancies were found to be negative U \cite{Anderson75} centers within these 
approximations. In these studies, however, the defect levels in the gap could not 
be determined unambiguously, owing to the well-known problem of underestimation 
of band gaps using Kohn-Sham eigenvalues \cite{Hybertsen86}. Later studies used 
hybrid functionals \cite{Gavartin06,Broqvist06,Xiong05} to calculate the defect 
levels of oxygen vacancies in hafnia. These functionals, which were constructed  
to fix the band gap underestimation problem, found that the defect was no longer 
a negative U center \cite{Broqvist06}. Recently, there has been a higher level 
theory study \cite{Choi09} using a combined DFT and GW approach on these defects. 
These authors found a negative U behavior for the oxygen vacancy. The GW part of 
the study in Ref. \cite{Choi09} was, however restricted to a 24-atom super cell.

In this letter, we report a new study of the quasiparticle excitations and charge 
transition levels of oxygen vacancies in monoclinic hafnia. To insure the 
accuracy of the results, we computed the charge transition levels via two 
physical paths and took into account an important electrostatic correction that 
was not included in the previous studies. The defects are found to be positive U 
centers. Further, we develop a methodology to understand the defect charging  
energy. We use a combined approach \cite{Hedestrom06,*Rinke09} based on DFT and 
GW formalism \cite{Hybertsen86}. This formalism corrects for the error incurred 
in calculating formation energy and charge transition levels within standard DFT.  
The calculations were done using large super cells with 96 atoms. Such large 
super cells are necessary to minimize any spurious defect-defect interactions 
from overlap of the defect state wave functions in neighboring super cells. 

\begin{figure}
\includegraphics[width=8.0cm]{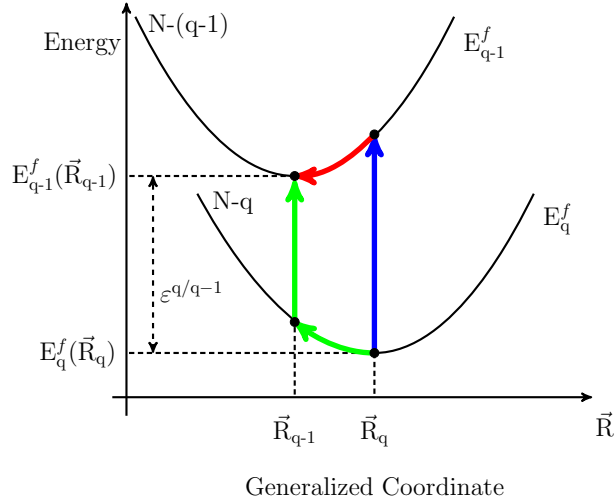}
\caption{Formation energies evaluated with $\mu = \text{E}_\text{v}$ vs. 
generalized coordinate, illustrating the terms in the DFT+GW formalism for the 
charge transition level $\varepsilon^{\text{q/q-1}}$}
\label{fig1}
\end{figure}

The formation energy of a defect in charge state q and at arbitrary ionic 
coordinates $\vec{\text{R}}$ and chemical potential $\mu$, $\text{E}_\text{q}^f(\vec{\text{R}})[\mu]$, can be expressed as 
\begin{equation}
\text{E}_\text{q}^f(\vec{\text{R}}) [\mu]= \text{E}_{\text{q}}(\vec{\text{R}}) - \text{E}_\text{ref} + \mu \text{q} 
\end{equation}
where $\text{E}_{\text{q}}(\vec{\text{R}})$ is the total energy of the system in 
charge state q and ionic coordinates $\vec{\text{R}}$, and $\text{E}_\text{ref}$ 
is the energy of a reference system with the same number of atoms as the charged 
system. We note that $\vec{\text{R}}$ is an arbitrary configuration which needs 
not be the equilibrium configuration of the charge state q which we denote as 
$\vec{\text{R}}_\text{q}$. Charge transition level, $\varepsilon^{\text{q/q-1}}$, 
is defined as the value of the chemical potential at which the charge state of 
the defect changes from q to q-1 (q/q-1). Conventionally, one defines the charge 
transition level from the valence band maximum, $\text{E}_\text{v}$. It is 
defined as the chemical potential $\mu$ at which the formation energies of the q 
and q-1 defects are equal and can be written in terms of formation energies as:
\begin{equation}
\varepsilon^{\text{q/q-1}} = \text{E}^f_\text{q-1}(\vec{\text{R}}_\text{q-1}) [\mu = \text{E}_\text{v}] - \text{E}^f_\text{q}(\vec{\text{R}}_\text{q}) [\mu = \text{E}_\text{v}] 
\label{eqn2}
\end{equation}
Within standard DFT, charge transition level is determined by calculating each of 
the formation energies in Eq. \ref{eqn2} in their respective equilibrium 
configurations as accurately as possible.  But, because of the band gap problem 
and self-interaction terms within standard DFT methods, significant errors may be 
introduced.  However, within the combined DFT and GW formalism, a charge 
transition level is written as: 
\begin{align}
\nonumber
\varepsilon^{\text{q/q-1}} =& [\text{E}^f_\text{q-1}(\vec{\text{R}}_\text{q-1}) - \text{E}^f_\text{q-1}(\vec{\text{R}}_\text{q})] + [\text{E}^f_\text{q-1}(\vec{\text{R}}_\text{q}) - \text{E}^f_\text{q}(\vec{\text{R}}_\text{q})] \\
\equiv& \text{E}_\text{relax} + \text{E}_\text{QP}
\label{eqn3}
\end{align}
by adding and subtracting the term 
$\text{E}^f_\text{q-1}(\vec{\text{R}}_\text{q})$ \cite{Hedestrom06,*Rinke09}. 
(All the formation energies in the above expression as well as throughout the 
rest of the letter are evaluated with $\mu = \text{E}_\text{v}$.) This 
reformulation allows us to combine terms to eliminate most of the errors 
mentioned above. Figure \ref{fig1} shows schematically the procedure for 
calculation of charge transition levels within the DFT+GW formalism. The first 
bracketed term on the right hand side of the Eq. \ref{eqn3} is a relaxation 
energy E$_\text{relax}$ (red line in Figure \ref{fig1}) and the second bracketed 
term is a quasiparticle excitation energy E$_\text{QP}$ (blue line in Figure \ref{fig1}).  
For the excitation energy, one uses the GW formalism \cite{Hybertsen86}; while for 
the relaxation energy, one uses DFT. This ensures an accurate calculation of the 
appropriate physical quantities and hence the charge transition level and 
relative formation energies.

It is worth noting that the charge transition level is a thermodynamic quantity 
and does not depend on the path in the formation energy-generalized coordinate 
space one takes to calculate it. In other words, the value of charge transition 
level remains unaffected when one adds and subtracts any formation energy to it. 
In particular, we can alternatively choose to add and subtract 
$\text{E}^f_\text{q}(\vec{\text{R}}_\text{q-1})$ in Eq. \ref{eqn2}. 
This would correspond to another path (the green line) in Figure \ref{fig1}. 
Calculating the charge transition level via multiple paths not only serves as a 
check for our calculations, but also gives some idea about the accuracy of the 
method. 

For the DFT part of our calculation we used an {\em ab initio} pseudopotential 
plane-wave method, as implemented in {\sc Quantum ESPRESSO} \cite{espresso}, with 
PBE \cite{Perdew96} exchange correlation functional. We used non-local 
pseudopotentials constructed using the Troullier Martins \cite{Troullier91} 
scheme with valence configurations $5\text{s}^25\text{p}^65\text{d}^26\text{s}^2$ 
and $2\text{s}^22\text{p}^4$ for Hf and O, respectively. The electronic wave 
functions were expanded in plane waves with cutoff energy of 250 Ry. The k-point 
sampling was restricted to the $\Gamma$ point in view of the large super cell 
used. Our calculated lattice parameters for monoclinic hafnia are in excellent 
agreement with experiment \cite{Wang92,*Adams91,*Stacy72} as well as with 
previous calculations \cite{Zheng07,*Kang03}. Using the PBE eigenvalues and 
eigenfunctions, the quasiparticle energies were calculated within the 
$\text{G}_0\text{W}_0$ approximation \cite{Hybertsen86} to the electron self-
energy as implemented in the BerkeleyGW package \cite{berkeleygw}. The static 
dielectric matrix was calculated with a 10 Ry energy cutoff in a plane wave basis 
and extended to finite frequencies within the generalized plasmon pole model 
\cite{Hybertsen86}. The band gap for bulk monoclinic hafnia is calculated to be 
6.0 eV, which is in good agreement with previous studies 5.45-5.9 
\cite{Choi09,Jiang10,*Gruning10} as well as experiments 5.7-5.9 
\cite{Sayan04,Bersch08}. 

There are two distinct types of oxygen vacancies (with different coordination) in 
monoclinic hafnia -- 3-fold coordinated (V$_{\text{O}_3}$) and 4-fold coordinated 
(V$_{\text{O}_4}$). We performed calculations for charge states q = 0, 1, 2 for 
both kind of vacancies, {\em i.e.} zero, one and two missing electrons from the 
vacancy.
\begin{figure}
\includegraphics[width=8.0cm]{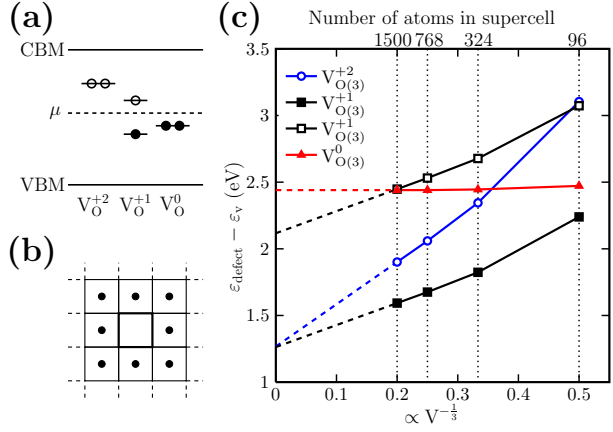}
\caption{(a) Schematic diagram showing both three and four fold coordinated 
oxygen vacancy induced defect levels in the band gap of hafnia for various charge 
states. (b) Periodic images of charged defects leading to a spurious 
electrostatic potential on the defect. (c) Effect of this spurious potential on 
eigenvalues. The plot shows cell size vs PBE eigenvalue with respect to the top 
of the valence band for vacancy induced defect states in V$_{\text{O}_3}$. The 
dashed line is the linear extrapolation to infinite super cell size.}
\label{fig2}
\end{figure}
\begin{table}
\squeezetable
\begin{tabular}{lcc|cc|cc|cc}
   \hline
   &  \multicolumn{4}{c|} {V$_{\text{O}_3}$} 
   & \multicolumn{4}{c} {V$_{\text{O}_4}$} \\
   \cline{2-9}
   & \multicolumn{2}{c|}{+2/+1} & \multicolumn{2}{c|} {+1/0} & \multicolumn{2}{c|} {+2/+1} & \multicolumn{2}{c} {+1/0} \\
   \cline{2-3} \cline{4-5} \cline{6-7} \cline{8-9}
   & P1 & P2 & P1 & P2 & P1 & P2 & P1 & P2 \\
   \hline
  E$_\text{relax}$ & --0.64 & \phantom{--}0.76 & --0.57 & \phantom{--}0.67 & --0.75 & \phantom{--}0.80 & --0.55 & \phantom{--}0.65 \\
  E$_\text{QP}$ & \phantom{--}3.30 & \phantom{--}1.69 & \phantom{--}3.93 & \phantom{--}2.88 & \phantom{--}3.04 & \phantom{--}1.33 & \phantom{--}3.46 & \phantom{--}2.50 \\
  $\varepsilon^{\text{q/q-1}}$  & \phantom{--}2.66 & \phantom{--}2.45 & \phantom{--}3.36 & \phantom{--}3.55 & \phantom{--}2.29 & \phantom{--}2.13 & \phantom{--}2.91 & \phantom{--}3.15 \\
  \hline
  Avg. $\varepsilon^{\text{q/q-1}}$ & \multicolumn{2}{c|} {2.56} & \multicolumn{2}{c|} {3.46} & \multicolumn{2}{c|} {2.21} & \multicolumn{2}{c} {3.03} \\
   \hline
\end{tabular}
\caption{Table showing the two components contributing to the charge transition 
levels for V$_{\text{O}_3}$ and V$_{\text{O}_4}$ calculated within the DFT+GW 
methodology. P1 (P2) refers to Path 1 (2). Row 1 shows the relaxation energy (E$_
\text{relax}$), Row 2 the quasiparticle energy (E$_\text{QP}$), Row 3 the charge 
transition level ($\varepsilon^{\text{q/q-1}}$) and Row 4 the value for $
\varepsilon^{\text{q/q-1}}$ averaged over both paths. All values are in eV.}
\label{table1}
\end{table}

Figure \ref{fig2} (a) shows a schematic of the oxygen vacancy induced 
quasiparticle defect levels in the band gap of hafnia. The defect levels for all 
the charge states lie deep in the gap. Table \ref{table1} shows the relaxation 
energies (E$_\text{relax}$), quasiparticle energies (E$_\text{QP}$) and charge 
transition levels ($\varepsilon^{\text{q/q-1}}$) calculated within DFT+GW 
approach. Despite the large differences in the quasiparticle energies required 
for the calculation in the two paths, the final charge transition levels via the 
two paths are always within 0.2 eV of one another. This gives us an error 
estimate of $\pm 0.1$ eV for the DFT+GW method for calculation of charge 
transition levels.

It is worth noting that quasiparticle energies in Table \ref{table1} include an 
electrostatic correction owing to the super cell geometry used. Within the 
standard DFT-only methodology of calculating charge transition levels, one 
corrects the total energies of charged defects for unphysical electrostatic terms 
from the charge on the image defects using Makov-Payne \cite{Makov95} like 
corrections. Within the DFT+GW method, the electrostatic error in 
the DFT eigenvalues in general needs to be accounted for. The origin of these electrostatic 
errors in the DFT eigenvalues is the spurious electrostatic potential from 
charged neighboring defects \cite{Lany10,*Freysoldt09} as shown schematically in 
Figure \ref{fig2} (b). To quantify these errors, we plotted the position of the 
Kohn-Sham defect level with respect to the valence band maximum for various super 
cell sizes as shown in Figure \ref{fig2} (c). These calculations were done using 
the {\sc SIESTA} code \cite{ordejon96siesta, *soler02siesta} with increasing 
super cell sizes - the largest of which contained 1499 atoms + defect. To get the 
infinite cell size value, we linearly extrapolated from the largest two super 
cell sizes. Figure \ref{fig2} (c) shows that even at 1500 atoms, the errors are 
large with respect to the converged answer. Table \ref{table2} shows the effect 
of electrostatic corrections on quasiparticle defect levels along the two paths 
for V$_{\text{O}_3}$. The two paths have different electrostatic corrections to 
the quasiparticle energies and that these corrections are substantial. Accounting 
for these errors is important to get a reliable value of charge transition level.  
\begin{table}
\begin{tabular}{lcc|cc}
   \hline
   & \multicolumn{2}{c|}{+2/+1} & \multicolumn{2}{c} {+1/0} \\
   \cline{2-3} \cline{4-5} 
   & P1 & P2 & P1 & P2  \\
   \hline
  E$_\text{QP}$ (Uncorrected) & 5.14 & 2.66 & 4.90 & 2.88 \\
  E$_\text{QP}$ (Corrected) &  3.30 & 1.69 & 3.93 & 2.88\\
  $\Delta \text{E}_\text{QP}$  & 1.84 & 0.97 & 0.97 & 0.00 \\
  \hline
  \end{tabular}
\caption{Table showing effect of spurious charged super cell electrostatic 
corrections on the quasiparticle defect levels for V$_{\text{O}_3}$ (referred 
to the top of the valence band). P1 (P2) refers to Path 1 (2). Row 1 shows the 
uncorrected quasiparticle energy, Row 2 the corrected quasiparticle energy, 
Row 3 the correction. All values are in eV.}
\label{table2}
\end{table}

Our quasiparticle defect levels and charge transition levels are in disagreement 
with previous DFT+GW calculations \cite{Choi09}. This disagreement is likely a 
consequence of their choice of a small super cell and, more importantly, neglect 
of electrostatic corrections. Reference \cite{Choi09}'s value of 
$\varepsilon^{+2/+1}$  for V$_{\text{O}_3}$ (V$_{\text{O}_4}$) is 4.00 eV (3.22 eV) 
and of $\varepsilon^{+1/0}$ is 3.10 eV (2.43 eV); these values are close to our 
corresponding {\em uncorrected} values. Our calculations also disagree with the 
hybrid functional value \cite{Broqvist06} of 3.7 eV for  $\varepsilon^{+2/+1}$ in 
V$_{\text{O}_3}$ and 4.0 eV for $\varepsilon^{+1/0}$ between V$^{+1}_{\text{O}_3}$ 
and V$^{0}_{\text{O}_4}$. (Our corresponding values are 2.56 eV and 3.32 eV 
respectively.) It is also worth noting that the Kohn-Sham eigenvalues calculated 
with hybrid functionals also disagree with our calculations. 
Also, in hybrid calculations, even though total energies were corrected 
for electrostatic errors, the eigenvalues were not. While this is not necessary 
for the charge transition level calculated as a total energy difference, it can lead 
to erroneous results whenever  eigenvalues are directly compared to single-particle excitation 
experiment.
\begin{figure}
\includegraphics[width=8.0cm]{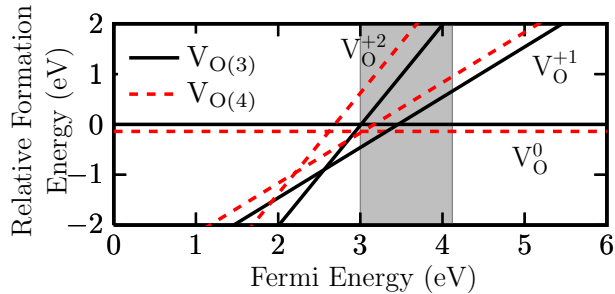}
\caption{Relative formation energy vs. chemical potential for charged oxygen vacancies in monoclinic hafnia.  The shaded region denoted the Si band gap. Thee formation energy of V$^0_{\text{O}_4}$ is lower than V$^0_{\text{O}_3}$ by 0.14 eV.}
\label{fig3}
\end{figure}

Figure \ref{fig3} shows the relative formation energy of various charge states 
for the oxygen vacancies as a function of the chemical potential in monoclinic 
hafnia based on the values in Table \ref{table1}. The formation energy in Figure 
\ref{fig3} is plotted with respect to the formation energy of V$^0_{\text{O}_3}$.  
As evident from Figure \ref{fig3}, V$^{+2}_{\text{O}_3}$, V$^{+1}_{\text{O}_3}$ 
and V$^0_{\text{O}_4}$  are the most stable defects in the system depending on 
the value of $\mu$. The experimentally relevant charge transition levels, 
$\varepsilon^{+2/+1}$ and $\varepsilon^{+1/0}$ would be 2.56 eV and 3.32 eV 
respectively. Not shown in the figure is the stability for 
V$^{-1}_\text{O}$/V$^{-2}_\text{O}$ defects. For larger values of the chemical 
potential, the V$^{-1}_\text{O}$/V$^{-2}_\text{O}$ defects may become more 
stable. Also, shown in Figure \ref{fig3} (in shaded grey), is the band gap and the expected band offset for Si ($\sim$3 eV) when placed next to hafnia 
\cite{Bersch08}. For p-doped Si next to hafnia, the system is expected to have 
V$^{+1}_{\text{O}_3}$ vacancies.

It is also noted from Figure \ref{fig3} that oxygen vacancies are positive U 
centers. U is defined as the energy of the reaction: $2 \text{V}^{+1} \rightarrow 
\text{V}^{2+} + \text{V}^{0}$. In terms of the charge transition levels, U for V$^{+1}$ can be written as:
\begin{align}
\text{U} &= \text{E}^f_{+2}(\vec{\text{R}}_{+2}) + \text{E}^f_{0}(\vec{\text{R}}_{0}) - 2 \text{E}^f_{+1}(\vec{\text{R}}_{+1}) \\
&= -\varepsilon^{+2/+1} + \varepsilon^{+1/0}
\nonumber
\label{eqn4}
\end{align}
Our calculated values of U for both vacancies (V$^{+1}_{\text{O}_3}$ and 
V$^{+1}_{\text{O}_4}$) are given in Table \ref{table3}. Further, U can be broken 
into two parts -- an electronic part (U$_\text{elec}$) and a relaxation part (U$_
\text{relax}$) as defined in the curly brackets in the Eq. 5 below:
\begin{align}
\text{U} = & \{\text{E}^{f}_{+2}(\vec{\text{R}}_{+1}) + \text{E}^{f}_{0}(\vec{\text{R}}_{+1}) - 2 \text{E}^{f}_{+1}(\vec{\text{R}}_{+1})\} + \\ 
\nonumber
&\{[\text{E}^{f}_{+2}(\vec{\text{R}}_{+2}) - \text{E}^{f}_{+2}(\vec{\text{R}}_{+1})] + [\text{E}^{f}_{0}(\vec{\text{R}}_{0}) - \text{E}^{f}_{0}(\vec{\text{R}}_{+1})]\} \\
\equiv &\text{U}_{\text{elec}} + \text{U}_{\text{relax}}
\nonumber
\label{eqn5}
\end{align}
This partitioning of U is instructive, because physically, U$_\text{elec}$ 
represents the quasiparticle gap of the system in the +1 charge state  keeping 
the structure fixed and U$_\text{relax}$ represents sum of structural relaxation energies. 
Physically, U$_\text{elec} \ge 0$ and U$_\text{relax} \le 0$. Table \ref{table3} 
shows our calculated values of U$_\text{relax}$ and U$_\text{elec}$. The reason 
for large relaxation energy is that, in the +1 and +2 charge state, the atoms 
nearest to the vacancy relax by up to 5-10\% of their bond lengths. This is in 
agreement with previous studies of relaxation around the vacancies 
\cite{Foster02,Zheng07,*Kang03}.
\begin{table}
\begin{tabular}{cccc}
   \hline
   & $\text{U}_\text{elec}$ & $\text{U}_\text{relax}$ & U$_\text{total}$ \\
   \hline
   {V$^{+1}_{\text{O}_3}$} & 2.24 & -1.33 & 0.90 \\
   {V$^{+1}_{\text{O}_4}$} & 2.13 & -1.35 & 0.81 \\
   \hline
\end{tabular}
\caption{U for V$^{+1}_{\text{O}_3}$ and V$^{+1}_{\text{O}_4}$ calculated using 
the charge transition levels from Table \ref{table1}. Also shown are the 
contributions to U from electronic and relaxation components.}
\label{table3}
\end{table}

In conclusion, we have reported quasiparticle energies and charge transition levels of 
oxygen vacancies in monoclinic hafnia. We find that V$^{+2}_{\text{O}_3}$, 
V$^{+1}_{\text{O}_3}$ and V$^{0}_{\text{O}_4}$ are the most stable oxygen 
vacancies in the system as the Fermi level spans the band gap. By calculating the 
charge transition levels via two paths in configuration-space we gain insight to 
the charge defect stability and highlighted the importance of electrostatic 
corrections in super cell defect calculations. Further, we developed an intuitive 
partitioning of the defect charging energy U into a quasiparticle gap and 
sum of relaxation energies. Contrary to some previous studies, which found negative 
U or U close to zero, the oxygen vacancies were found to be positive U centers.

M.J. would like to thank Brad Malone and Dr. Georgy Samsonidze for fruitful 
discussions. This work was supported by National Science Foundation Grant No. 
DMR07-05941, the U.S. Department of Energy under Contract No. DE-AC02-05CH11231 
and DE-SC0001878. Computational resources have been provided by NSF through 
TeraGrid resources at NICS. M.J. was supported by the DOE. Part of the modeling 
and simulations were carried out with electronic structure and quasiparticle 
codes developed under NSF support. 

%

\end{document}